\documentclass[aip, jmp, amsmath,amssymb, reprint]{revtex4-1}
\usepackage[version=3]{mhchem} % Formula subscripts using \ce{}

\usepackage{times,amsmath}
\usepackage{epsfig}
\usepackage{color}
\usepackage{longtable}
\usepackage{graphicx}% Include figure files
\usepackage{dcolumn}% Align table columns on decimal point
\usepackage{bm}% bold math
\usepackage{tabularx}% bold math
\usepackage{hyperref}
\usepackage{multirow}
\hypersetup{citecolor=black, filecolor= black, linkcolor= black, urlcolor= black}

\begin{document}
\preprint{AIP/123-QED}

\title{Predicting Polymeric Crystal Structures by Evolutionary Algorithms}
\thanks{These corresponding authors contributed equally to this work}
%\thanks{These corresponding authors contributed equally to this work: qiang.zhu@stonybrook.edu and vinit.sharma.mlsu@gmail.com}

\author{Qiang Zhu}
\email{qiang.zhu@stonybrook.edu}
\affiliation{Department of Geosciences, Stony Brook University, Center for Materials by Design, Institute for Advanced Computational Science, Stony Brook University, NY 11794, USA}
\author{Vinit Sharma}
\email{vinit.sharma.mlsu@gmail.com}
\affiliation{Materials Science and Engineering, Institute of Materials Science, University of Connecticut, Storrs, 06626, CT, USA}
\author{Artem R. Oganov}
\affiliation{Department of Geosciences, Stony Brook University, Center for Materials by Design, Institute for Advanced Computational Science, Stony Brook University, NY 11794, USA}
\affiliation{Department of Problems of Physics and Energetics, Moscow Institute of Physics and Technology, 9 Institutskiy lane, Dolgoprudny city, Moscow Region, 141700, Russia}
\affiliation{Department of Materials Science and Engineering, Northwestern Polytechnical University, Xi'an,710072, China}
\author{Rampi Ramprasad}
\affiliation{Materials Science and Engineering, Institute of Materials Science, University of Connecticut, Storrs, 06626, CT, USA}
\date{\today}


\begin{abstract}
The recently developed evolutionary algorithm USPEX proved to be a tool that enables accurate and reliable prediction of structures for a given chemical composition. 
Here we extend this method to predict the crystal structure of polymers by performing constrained evolutionary search, where each monomeric unit is treated as one or several building blocks with fixed connectivity.
This greatly reduces the search space and allows the initial structure generation with different sequences and packings using these blocks. 
The new constrained evolutionary algorithm is successfully tested and validated on a diverse range of experimentally known polymers, 
namely polyethylene (PE), polyacetylene (PA), poly(glycolic acid) (PGA), poly(vinyl chloride) (PVC),
 poly(oxymethylene) (POM), poly(phenylene oxide) (PPO), and poly (p-phenylene sulfide) (PPS).  
By fixing the orientation of polymeric chains, this method can be further extended to predict all polymorphs of poly(vinylidene fluoride) (PVDF), and  the complex linear polymer crystals, such as nylon-6 and cellulose.
The excellent agreement between predicted crystal structures and experimentally known structures assures a major role of this approach in the efficient design of the future polymeric materials.
\end{abstract}

\maketitle

\section{Introduction}

Recent methodological developments have made it possible to predict crystal structures of inorganic solids from only a knowledge of their composition. \cite{Woodley-NM-2008, Oganov-ACR-2011}  
Tools such as evolutionary algorithm USPEX (Universal Structure Predictor: Evolutionary Xtallography) enable accurate and reliable prediction of structures for a given chemical composition at given pressure--temperature conditions. 
In particular, for inorganic crystals, USPEX has been successfully used to predict the stable structures of a wide range of novel materials at normal, and extreme pressure conditions. \cite{Oganov-ACR-2011}

Even after decades, the comment by Maddox \cite{Maddox-1988} ``One of the continuing scandals in the physical sciences is that it remains impossible to predict the structure of even the simplest crystalline solids from a knowledge of their composition" remains valid. 
The ability to predict the key physical and chemical properties of polymers from their molecular structure can be of great value in the design of polymers for numerous technological applications such as capacitive energy storage, transistors and photovoltaic devices.\cite{ma2024057,Chu21072006,cm102419z,ja202755x,Wang2014979} 

In organic crystals, if the molecules already satisfy the bonding requirements, they form molecular crystals. 
On the other hand, unsaturated molecules tend to polymerize into long-chain molecules in the liquid state without long-range order. 
Under appropriate conditions, polymers might crystallize - that is, polymeric chains can develop order and arrange themselves periodically.  
One can construct a tremendous variety of polymeric structures based on the same monomeric blocks, all with different stability and properties.
%The evolution of model systems of length scale of about 100 \AA ~for times up to a few tens of nano seconds. 
%Hence, it requires not only accurate interatomic potentials but also enormous computing time \cite{Wang2014979}.
 
Past attempts for prediction of 3D packing of organic molecules were based on different energy minimization methods.\cite{Blindtest-2000,Blindtest-2002,Blindtest-2005,Blindtest-2009,Blindtest-2011}.  
In our recent work, we proved that evolutionary algorithms (EA) can efficiently solve this problem. \cite{Zhu-Acta-2012}
To the best of our knowledge, so far no attempt has been made to predict the structure of crystalline polymers that possess maximal stability or optimal physical properties.
In this paper, we present a powerful EA-based technique, and its power is demonstrated by the successful identification of various experimentally known polymers.

\section{Evolutionary Algorithms}
The USPEX code has been successfully applied to various classes of systems (bulk crystals \cite{Oganov-JCP-2006, Oganov-ACR-2011}, nanoclusters \cite{Lyakhov-CPC-2013}, 2D crystals \cite{Zhou-arxiv-2013} and surfaces \cite{Zhu-PRB-2013}).
Extending the range of its applicability, we proposed a new \emph{constrained global optimization} method to predict the packing of molecular crystals. \cite{Zhu-Acta-2012} 
A similar concept can be applied to polymers as well. 
If we start to search for the global energy minimum with randomly generated structures (according to the given chemical formula), it is very likely that most of the time will be spent on exploring many disordered structures characterized by irrelevant structural motifs. 
More importantly, the desired polymeric crystals are usually not the thermodynamic equilibrium in the given chemical system. 
The truly interesting searching target is actually the optimum sequence of monomers, and 3D-packing of the pre-formed polymeric chains. 
This problem can be solved by \emph{constrained global optimization} - finding the most stable packing of monomers with fixed bond connectivity. 
It requires whole motifs rather than individual atoms to be considered as the minimum building blocks in our search.
This strategy does not only make the global optimization meaningful, but at the same time simplifies it, leading to a drastic reduction of the number of variables in the search space.  

\begin{figure}
\epsfig{file=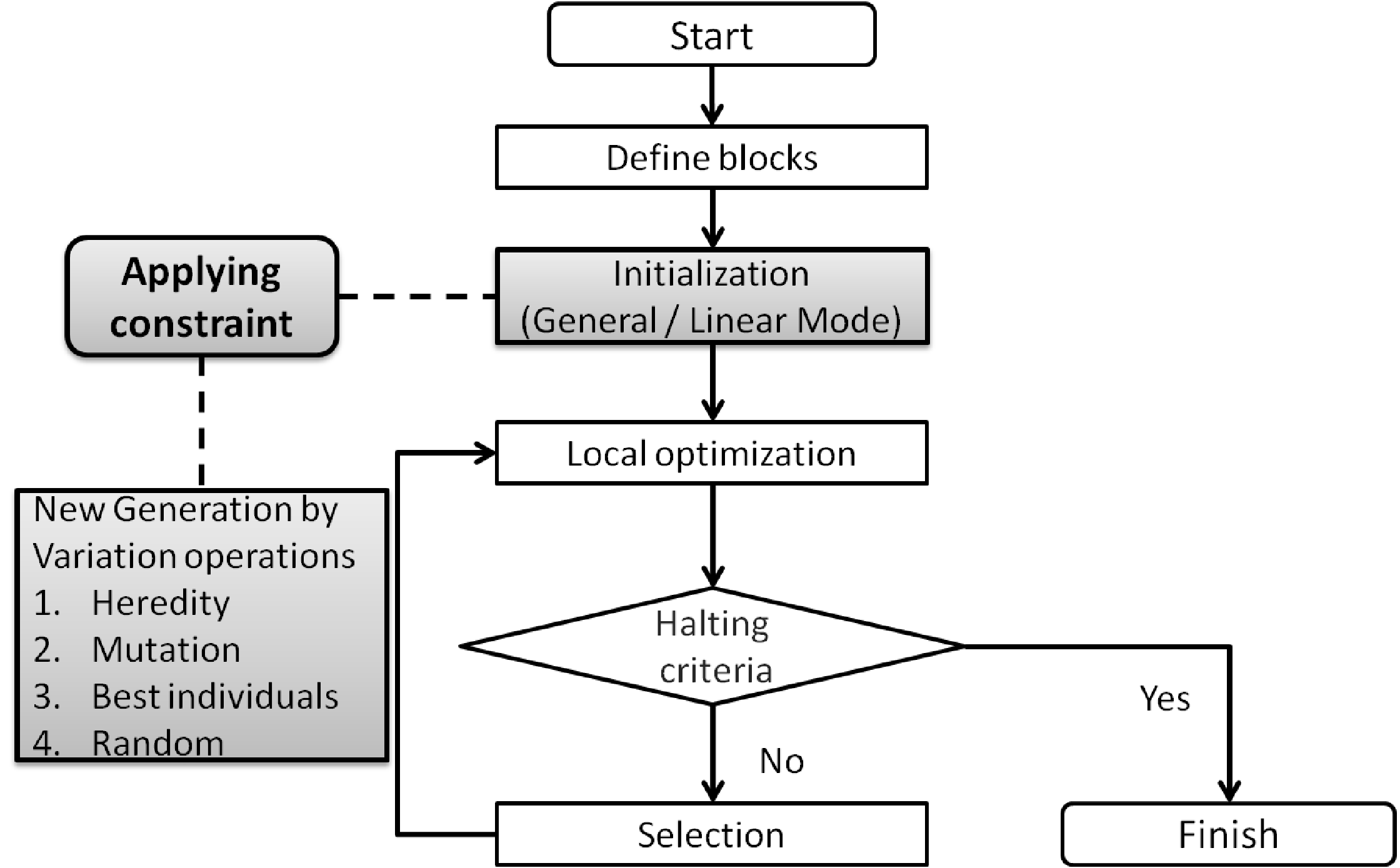, width=0.45\textwidth}
\caption{\label{flowchart} Flowchart of the constrained evolutionary algorithm. The key feature is that all the generated structures before local optimization are constructed based on the pre-specified building blocks (highlighted in grey box).}
\end{figure}

In the context of EA, our procedure is as shown in Fig \ref{flowchart}.

\begin{figure}
\epsfig{file=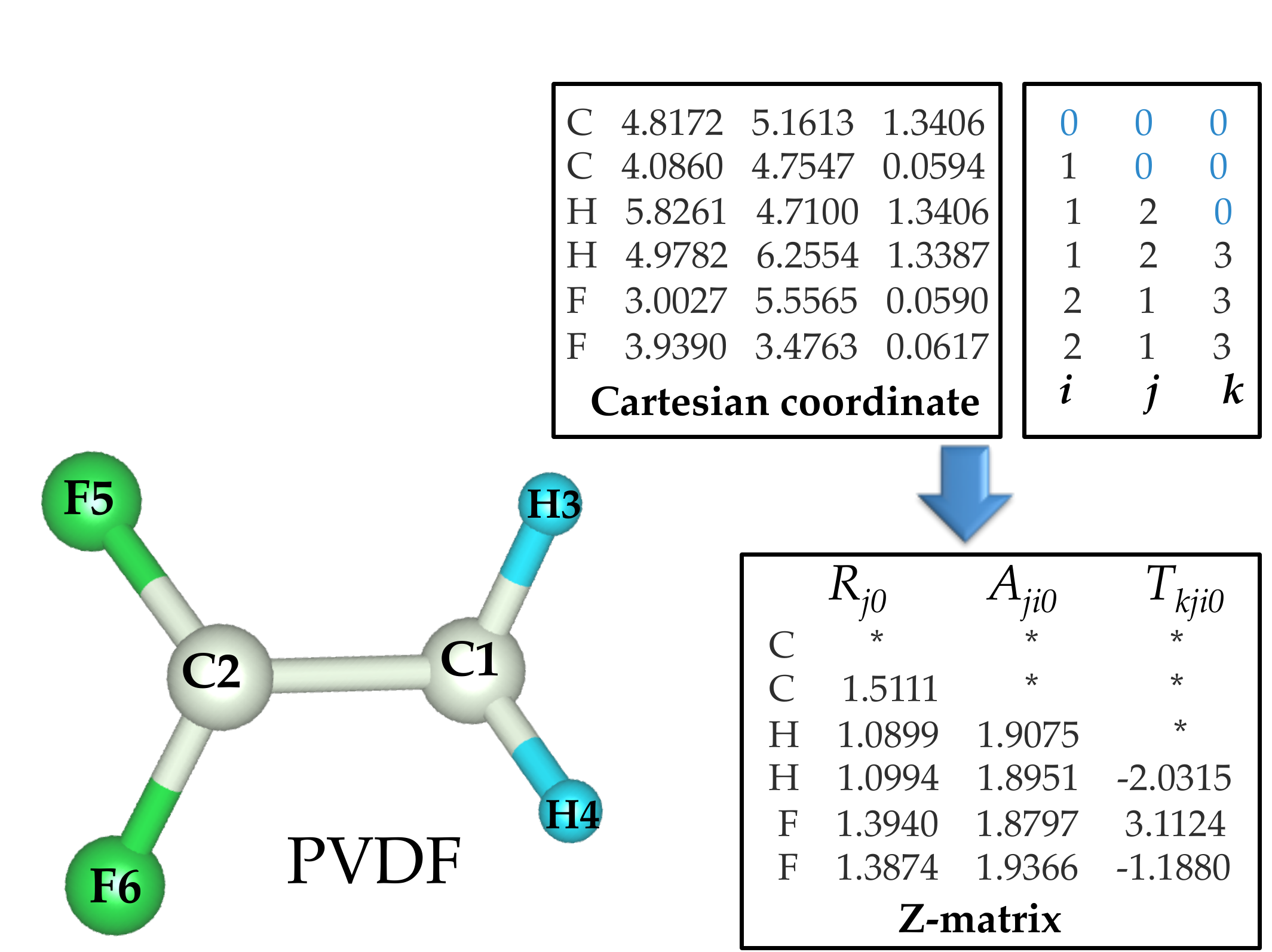, width=0.4\textwidth}
\caption{\label{Z-Matrix} Transformation from Cartesian coordinates to Z-matrix representation. The $i-j-k$ table specifies the topological relations in the Z-matrix style. $R$ is distance from atom 0 to atom number $i$, $A$ is the angle made by the present atom with atoms $j$ and $i$, while $T$ is the torsional angle made by the present atom with atoms $k$, $j$, and $i$.}
\end{figure}

(1) {\bf Define Blocks}. Since the monomeric units can be made of one or multiple types of blocks, we represent them by using Z-matrix, which has been widely used to describe the molecular structure in organic chemistry. 
For the atom in a given molecule, its bond connectivity can be defined by internal coordinates (i.e., the bond length, bond angle and torsional angle). 
As shown in Fig \ref{Z-Matrix}, Cartesian coordinates can be transformed to Z-matrix representation according to the constraints by those internal coordinates. 
In Z-matrix, the top three atoms lack some constraints, since there are no reference atoms to define their internal coordinates. 
The 6 missing components in the Z-matrix correspond the 3 translational and 3 rotational variables in 3D space. 
From now on, we treat each block as a rigid body, and construct the crystal structures by varying only those 6 variables in each Z-matrix. 

(2) {\bf Initialization}. 
At the beginning of EA, the structures for the first generation are randomly produced. 
A fully random initialization is a poor choice for large systems, as it always leads to nearly identical glassy structures that have similar (high) energies and low degree of order \cite{Lyakhov-CPC-2013}. 
From such starting conditions, it is difficult to obtain ordered crystalline states. 
To achieve both high structural diversity and quality, a better way is to create symmetric structures for the initial population. 
%There are two ways to construct the polymer crystals depending on the availability of knowledge about the polymer.

If the polymer consists of multiple types of blocks and the organization of them is unclear, we need to navigate all the possibilities. 
That is, we treat each block independently and generate structures with each block randomly located in the unit cell, just as we did for the prediction of molecular crystals \cite{Zhu-Acta-2012}. 
We firstly generate a random symmetric structure with the geometric centers of each block being located at general or special Wyckoff positions in a 3D primitive Lattice ($h$), by using a special random symmetric algorithm \cite{Lyakhov-CPC-2013}. 
For each Wyckoff site, the first molecular block ($R$) is built around it with random orientation, and the replica blocks ($R'$) can be obtained by symmetry operations, which are a combination of point group (P) and translations (T) operations,
\begin{equation}
R' = R\cdot{\textrm P} + {\textrm T}
\end{equation}

\begin{figure}
\epsfig{file=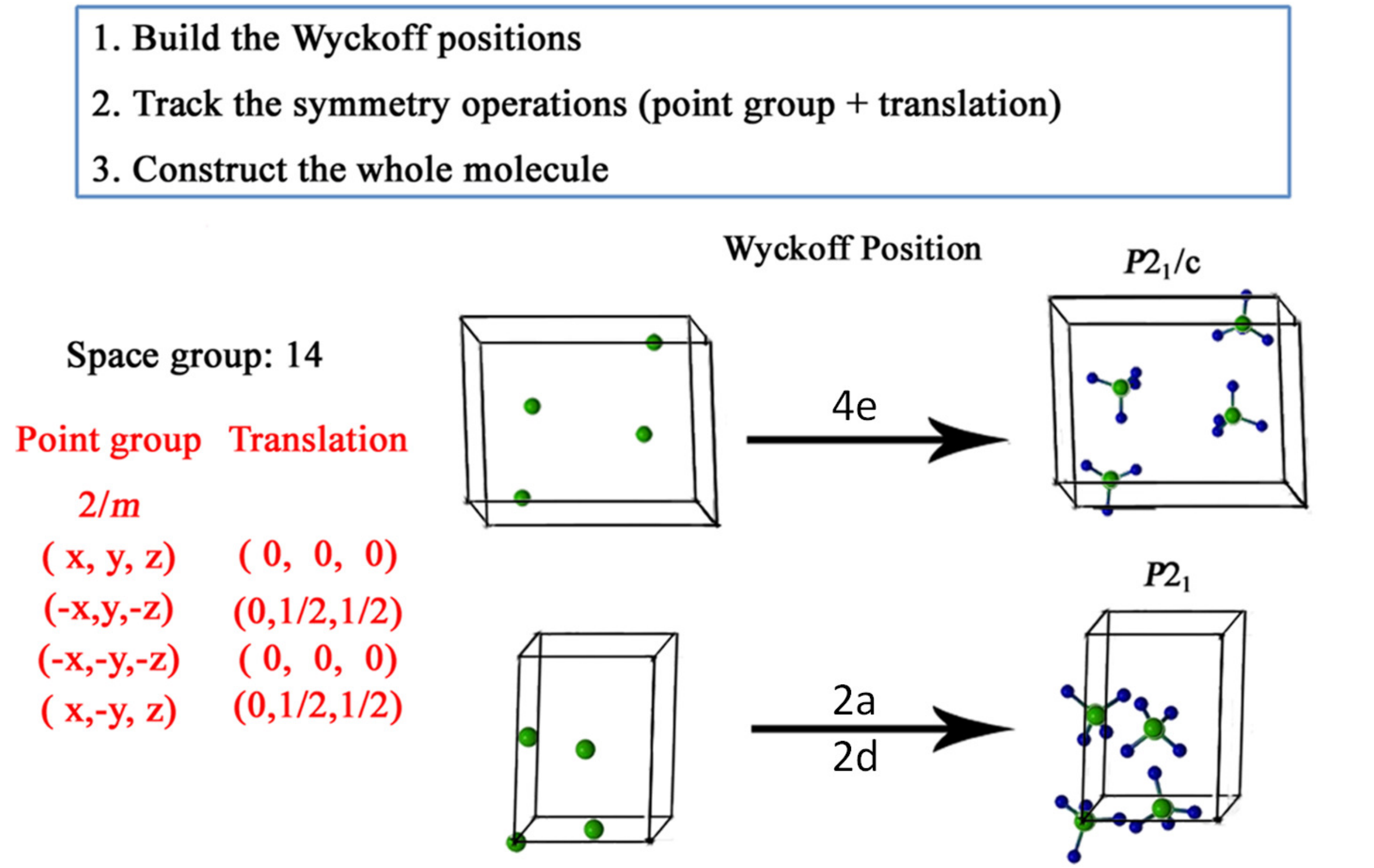, width=0.5\textwidth}
\caption{\label{Symmetry} Illustration of generating a random symmetric structure with 4 molecules per cell. For a given space group randomly assigned by the program (in this case, $P$$2_{1}$/$c$), the Bravais cell is generated, and molecular center is placed onto a random position (in this case, the general position 4e or 2a+2d). Molecules are then built at the Wyckoff sites according to their connectivity and with their orientations obeying space group symmetry operations. If the special Wyckoff sites are occupied, molecular geometry often breaks space group symmetry, leading to a subgroup. For clarity of the figure, molecules occupying positions at the corners and faces of the unit cell are shown only once. }
\end{figure}

This whole scheme is illustrated in Fig \ref{Symmetry}, which exactly works for molecules occupying the general Wyckoff site. 
If the special Wyckoff sites are involved, the generated structures are likely to have lower symmetry: if the molecular block itself has low symmetry, 
there will be symmetry breaking, leading to a subgroup symmetry, and we allow this. 

\begin{figure}
\epsfig{file=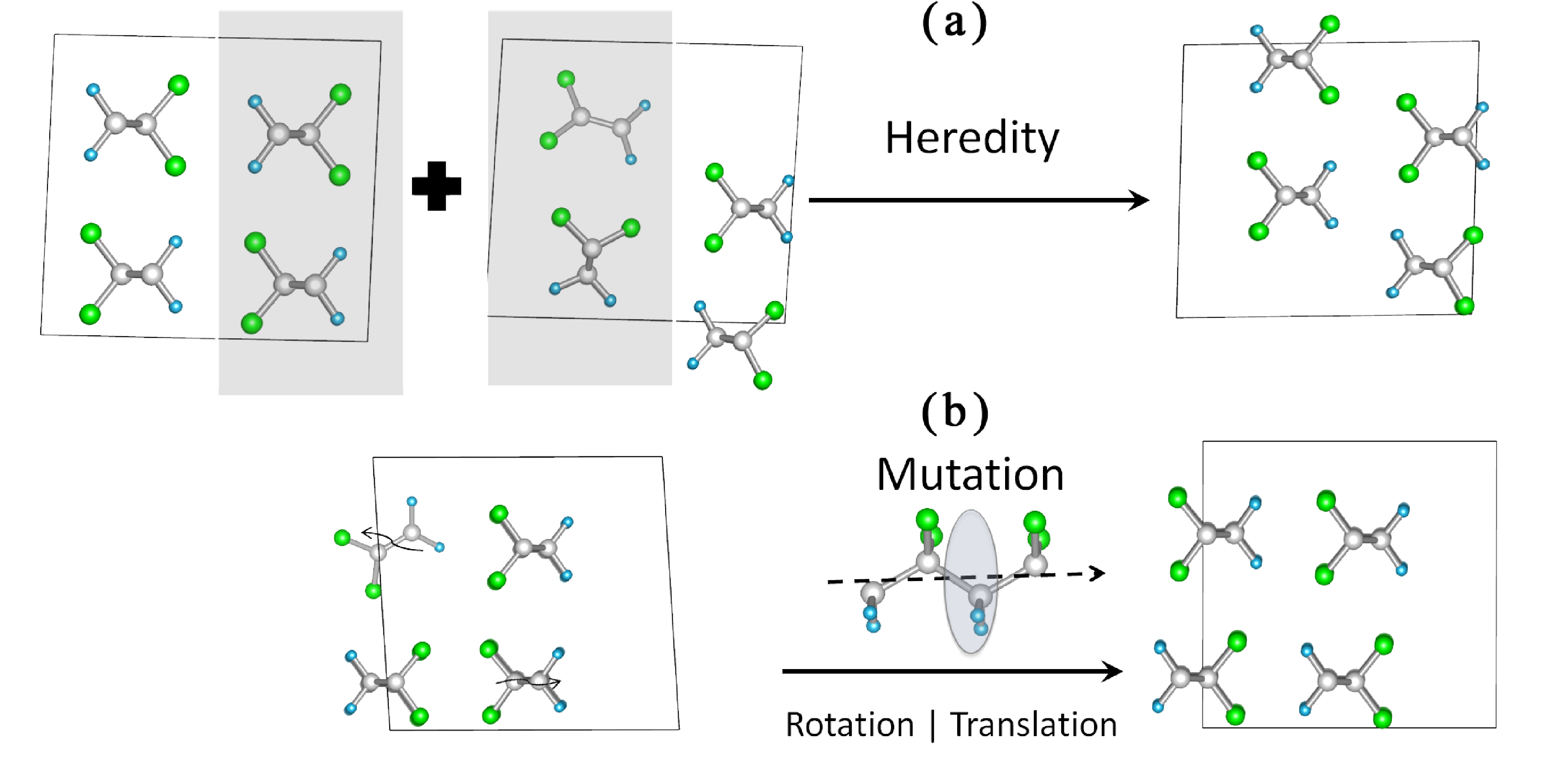, width=0.5\textwidth}
\caption{\label{variation} Variation operators in EA. (a) heredity; (b) mutation.}
\end{figure}

(3) {\bf Local optimization}. Structural relaxation is done stepwise from low to high precision, as described in Ref. \cite{Oganov-JCP-2006}, to achieve greater efficiency.

(4) {\bf Selection}. At the end of each generation, all relaxed structures in the generation are compared using their fingerprints \cite{Zhu-Acta-2012} and all non-identical structures are ranked by their (free) energies.

(5) {\bf Variation Operators}. Child structures (new generation) are produced from parent structures (old generation) using two general types of variation operators: heredity and several kinds of mutations \cite{Oganov-JCP-2006}. 
Heredity, mating two different parent structures, establishes communications between good candidate solutions. 
Mutations are aimed at introducing new features into the population and probing the neighborhood of the already found low-energy structures by strongly perturbing them. 
Different from those in atomic crystals, these variation operators act on the geometric centers of the molecules and their orientations. 
It is very important to note that symmetric initialization scheme might favor only the symmetric structures. 
Yet, our variation operators can break symmetry to make sure the even asymmetric ground states will not be missed.

{\it Heredity}: This is a basic variation operator in EA. 
It cuts planar slices from two selected individuals and combines them to produce a child structure. 
During this process, each molecule is represented by its geometric center and orientation (Fig \ref{variation}a),
, and the entirety of molecules is retained during the operation of heredity (as well as of all the other variation operators).

{\it Rotational mutation}: A certain number of randomly selected molecules are rotated by random angles. 
The rotational axes can be obtained by the eigendecomposition of the inertia tensor matrix of the given molecule (Fig \ref{variation}b).
\begin{equation}
[I] = [Q][\Delta][Q^T].
\end{equation}

The inertia tensor can be computed by
\begin{equation}
I = \sum_{i=1}^{N}{r_i^{2}},
\end{equation}

where $r$ is the corresponding distance from each atom to the geometric center. 

The columns of the rotation matrix [Q] define the directions of the principal axes of the body, and [$\Delta$] is a diagonal matrix,
\begin{equation}
[\Delta] = 
\begin{pmatrix}
I_1 & 0 & 0\\
0 & I_2 & 0\\
0 & 0 & I_3\\
\end{pmatrix}
\end{equation}
where, $I_1$, $I_2$ and $I_3$ are called principal moments of inertia, determining which direction in [Q] is easier to rotate. 
For the polymeric chain, $I_1$ is usually significantly larger than $I_2$ and $I_3$. 
And the first direction in [Q] nearly coincides with the direction of the polymers chain.

{\it Translational mutation}: All the centers of molecules are displaced in random directions, the displacement magnitude for molecule \emph{i} being picked from a zero-mean Gaussian distribution with $\sigma$ defined as:
\begin{equation}
\sigma_i = \sigma_{\rm max} \frac{\Pi_{\rm max}-\Pi_i}{\Pi_{\rm max}-\Pi_{\rm min}},
\end{equation}
where $\Pi$ is the local degree of order of the molecule \cite{Zhu-Acta-2012} and $\sigma_{\rm max}$ is the order of a typical intermolecular distance. 
We calculate the $\Pi$ of each molecule's geometric center from its fingerprint \cite{Oganov-JCP-2009}. 
Thus molecules with more ordered environment are perturbed less than molecules with less ordered environment.

{\it Softmutation}: This operator involves atomic displacements along the softest mode eigenvectors, or a random linear combination of softest eigenvectors \cite{Lyakhov-CPC-2010}. 
For molecular crystals, it becomes a hybrid operator, combining rotational and coordinate mutations. 
In this case, the eigenvectors are calculated first, and then projected onto translational and rotational degrees of freedom of each molecule and the resulting changes of molecular positions and orientations are applied preserving rigidity of the fixed intra-molecular degrees of freedom.
For the rapid calculation of vibrational modes, here we use the approach of Ref. \cite{Lyakhov-CPC-2010}.

{\it Addition of random structures}: Although the searching space has been effectively decreased by applying the geometry constraints, we are still facing a high dimensional configuration space. 
A general challenge for EA (and many other global optimization methods) is how to avoid getting stuck in a local minimum when dealing with multidimensional spaces - in other words, avoiding decrease of population diversity during evolution.
A key to maintain the population diversity is to add new blood. 
Therefore, we produce some fraction (usually 15\% - 30\%) of each generation using the random symmetric algorithm described above. Fingerprint niching also helps to retain the diversity \cite{Lyakhov-CPC-2010}.

(6) {\bf Halting and post-processing}. After the lowest-energy structure is unchanged for a certain number of generations, the calculation automatically stops and the lowest-energy structures found in USPEX are then carefully relaxed with higher precision: the all-electron projector-augmented wave (PAW) method \cite{PAW}, as implemented in the VASP code \cite{vasp}, at the level of generalized gradient approximation (GGA-PBE functional) \cite{GGA} and van der Waals (vdW) dispersion-corrected GGA (PBE-D \cite{DFT+D} and PBE-TS \cite{PRL-TS}, the latter also being used in USPEX structural relaxation). 
We used the plane wave kinetic energy cutoff of 550 eV and the Brillouin zone was sampled with a resolution of 2$\pi$ $\times$ 0.07 \AA$^{-1}$, which showed excellent convergences of the energy differences, stress tensors and structural parameters. 

\section{Prediction of the Crystal Structures of Simple Polymers}
The new constrained EA in combination of state-of-the-art quantum-mechanical computational methods successfully predict the crystal structures, lattice parameters and densities for several polymers composed of simple monomeric units, namely polyethylene (PE), polyacetylene (PA), poly(glutamic acid) (PGA), poly(vinyl chloride) (PVC), poly(oxymethylene) (POM), poly(phenylene oxide) (PPO), and poly (p-phenylene sulfide) (PPS).
We note that while past computational work starts at the experimentally known structural parameters and space group, no such assumptions were made in the present study. 
The symmetry and other structural parameters are the outcome of the search process. 
As the new constrained EA significantly reduces the search space, the correct structures of these known polymers are found in the first few generations.
Below we discuss these systems and their structures.

\begin{figure}
\epsfig{file=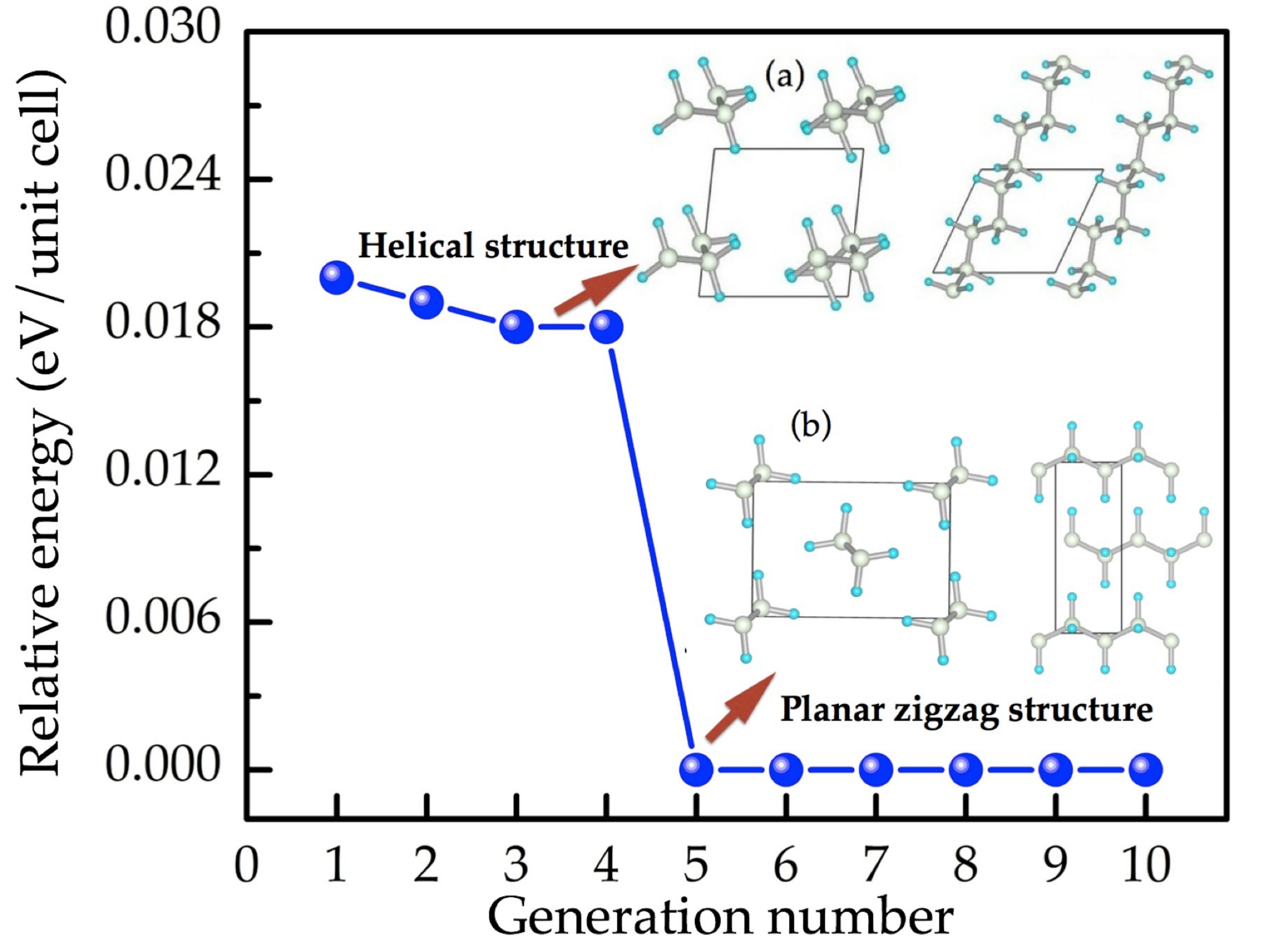, width=0.5\textwidth}
\caption{\label{Fig7-PE} The evolution of lowest-energy structure as a function of generation number for PE. The predicted stable structures (top and side view) of (a) The helical structure identified in a distinct local minimum for PE and (b) Planar zigzag structure are also shown. Grey circles represent carbon atoms and cyan circles hydrogen atoms. Grey and cyan spheres represent the carbon and hydrogen atoms, respectively.}
\end{figure}

{\bf PE}:
We start with the simplest polymer, polyethylene ([-CH$_2$-CH$_2$-]$_n$).
In Fig \ref{Fig7-PE}, the evolution of lowest-energy structure as a function of generation number for PE is plotted.
The carbon backbone of the equilibrium structure has a planar all-trans zigzag structure.
Its space group (\emph{Pnma}), equilibrium geometry and density was found to be in agreement with available experimental measurements and previous \textit{ab initio} results \cite{Avitabile17,Miller-1960,CSLiu}.
It is evident from the Fig \ref{Fig7-PE} that the new constrained evolutionary search scheme explored the configurational space in such a efficient manner that our algorithm is capable enough of finding the meta-stable phases in first few generations (\textless 5).
Our EA based structural search also identifies that in addition to the global minima, the total energy also has a distinct local minimum for the helical structure as shown in Fig \ref{Fig7-PE}, which is 0.018 eV/unit cell higher in energy than the ground state.

\begin{figure}
\epsfig{file=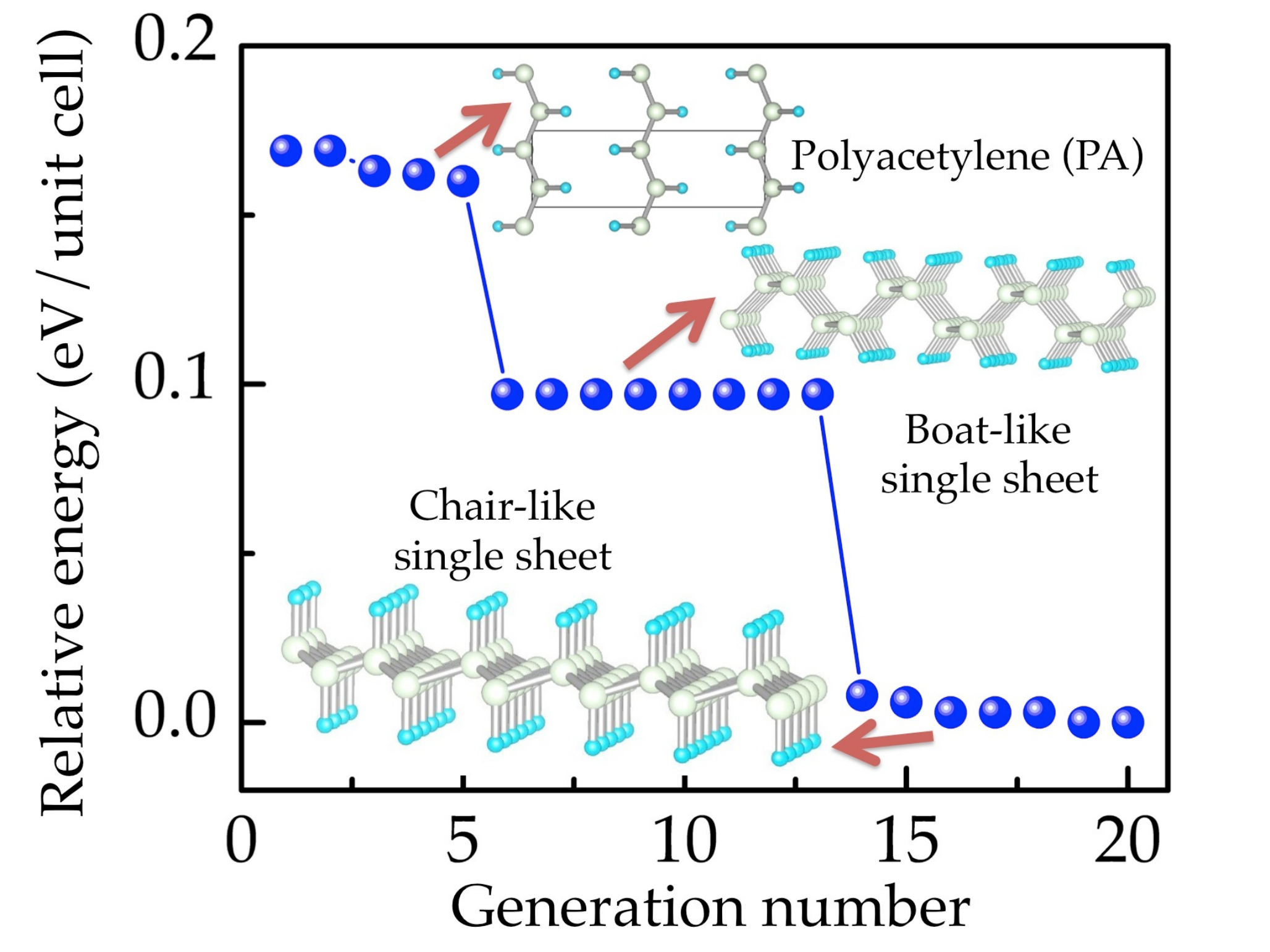, width=0.5\textwidth}
\caption{\label{Fig8-C2H2} The evolution of lowest-energy structure as a function of generation number for PA having repeating unit [-C$_2$H$_2$-]$_n$. Along with experimentally known structure two other stable boat-like and chair-like structures are also shown. Grey and cyan spheres represent the carbon and hydrogen atoms, respectively.}
\end{figure}

{\bf PA}:
The 3D geometry of the crystalline PA polymer with the repeat unit [-C$_2$H$_2$-]$_n$ has been a question of debate.
To resolve this issue, attempts have been made by various experiments \cite{Zhu-EXPT} as well as computations \cite{Wilson,Wen-2011}.
The two proposed structures with space groups $P$2$_1$/$a$ and $P$2$_1$/$n$ have a slight difference in the orientation of double bonds of adjacent chains.
In the case of $P$2$_1$/$a$, the double bonds of adjacent chains are in-phase while in the case of $P$2$_1$/$n$, they are out-of-phase.
Along the chain axis, translating alternate chains of the $P$2$_1$/$a$ structure by $c$/2 results in the $P$2$_1$/$n$ structure.
Our calculations predict that the structure where double bonds of adjacent chains are in-phase ($P$2$_1$/$a$) is more stable.
The evolution of lowest-energy structure as a function of generation number for PA is shown in Fig \ref{Fig8-C2H2}.
Calculated lattice parameters are in agreement with experimental values (Fig \ref{results}).
Moreover, it is worth mentioning that in this search we also obtained isomeric two-dimensional sheets of graphane with CH stoichiometry (shown in Fig \ref{Fig8-C2H2}), which are even more stable than both benzene (C$_6$H$_6$) and PA (C$_2$H$_2$)$_n$ \cite{Zhu-EXPT}.
The boat-like and chair like structures were also identified with a previous version of the USPEX method \cite{Wen-2011}.

\begin{table}
\caption{The predicted lattice parameters and density for all considered polymers, namely PE, PA, PPO, POM, PPS, PGA and PVC. For comparison available experimental and computations results at the same level of theory are also listed. All considered polymers have four formula units (f.u.) per unit cell (\textit{Z}=4), except PA has \textit{Z}=2.}
\label{results}
\begin{tabular}{clccccc}
\hline
\hline
Polymer & Method &\emph{a}({\AA}) &\emph{b}({\AA}) &\emph{c}({\AA}) &$\beta$($^\circ$) &Density \\
 &~&~&~&~&~& (g/cm$^3$)                                        \\
\hline
 \multirow{4}{*}{\begin{tabular}{c}PE \\ {[-CH$_2$-]$_n$} \\ (\emph{Pnma})   \end{tabular}}
 &Expt.\cite{Miller-1960}    &7.12 &4.85 &2.55 &~  &0.997  \\
 &PBE\cite{CSLiu}                                     &8.20 &5.60 &2.55 &~  &0.796  \\
 &PBE-TS\cite{CSLiu}                                  &7.01 &4.76 &2.56 &~  &1.091   \\
 &PBE-TS$^*$         &7.02 &4.76 &2.56 &~  &1.091    \\
\hline
\multirow{4}{*}{\begin{tabular}{c}PA \\ {[-C$_2$H$_2$-]$_n$} \\ (\emph{P}2$_1$/\emph{n})  \end{tabular}}
 &Expt.\cite{Miller-1960}               &4.24 &7.32 &2.46 &91-94   &1.130      \\
  &PBE\cite{CSLiu}                                     &5.00 &7.74 &2.46 &90.3  &0.908 \\
 &PBE-TS\cite{CSLiu}                                  &4.01 &7.19 &2.46 &90.3  &1.22  \\
 &PBE-TS$^*$               &4.00 &7.22 &2.46 &90.6  &1.191   \\
 \hline
  \multirow{4}{*}{\begin{tabular}{c}PGA \\ {[-C$_2$H$_2$O$_2$-]$_n$} \\ (\emph{Pcmn})  \end{tabular}}
 &Expt.\cite{Miller-1960}               &5.22 &6.19 &7.02 &~  &1.700  \\
 &PBE\cite{CSLiu}                                     &5.07 &5.58 &6.96 &~  &1.958 \\
 &PBE-TS\cite{CSLiu}                                  &5.09 &6.11 &7.03 &~  &1.763 \\
  &PBE-TS$^*$              &5.13 &6.09 &7.01 &~  &1.79  \\
 \hline
 \multirow{4}{*}{\begin{tabular}{c}PVC \\ {[-CH$_2$-CHCl-]$_n$} \\ (\emph{Pbcm})  \end{tabular}}
 &Expt.\cite{Miller-1960}                         &10.24 &5.24 &5.08 &~  &1.523      \\
 &PBE\cite{CSLiu}                                     &10.45 &5.50 &5.05 &~  &1.430  \\
 &PBE-TS\cite{CSLiu}                                  &10.11 &5.15 &5.08 &~  &1.540 \\
  &PBE-TS$^*$             &10.14 &5.16 &5.08 &~  &1.530    \\
\hline
 \multirow{4}{*}{\begin{tabular}{c}POM \\ {[-CH$_2$-O-]$_n$} \\ (\emph{P}2$_1$2$_1$2$_1$)  \end{tabular}}
 &Expt.\cite{Carazzolo36,Miller-1960}             &4.77 &7.65 &3.56 &~   &0.922        \\
 &PBE\cite{CSLiu}                                     &5.40 &8.37 &3.63 &~  &0.730   \\
 &PBE-TS\cite{CSLiu}                                  &4.59 &7.72 &3.57 &~  &0.947  \\
  &PBE-TS$^*$              &4.55 &7.75 &3.59 &~  &0.95  \\
 \hline
 \multirow{4}{*}{\begin{tabular}{c}PPO \\ {[-C$_6$H$_4$O-]$_n$} \\ (\emph{Pbcn})  \end{tabular}}
 &Expt.\cite{Miller-1960}                &8.07 &5.54 &9.72 &~   &1.408      \\
 &PBE\cite{CSLiu}                                    &8.42 &5.88 &9.85 &~  &1.254   \\
 &PBE-TS\cite{CSLiu}                                 &8.04 &5.37 &9.75 &~  &1.453 \\
  &PBE-TS$^*$                &8.02 &5.36 &9.75 &~  &1.491   \\
 \hline
 \multirow{4}{*}{\begin{tabular}{c}PPS \\ {[-C$_6$H$_4$S-]$_n$} \\ (\emph{P}\emph{bcn})  \end{tabular}}
 &Expt.\cite{Miller-1960}                         &8.67 &5.61 &10.26 &~  &1.440       \\
  &PBE\cite{CSLiu}                                     &8.85 &5.73 &10.26 &~  &1.381  \\
  &PBE-TS\cite{CSLiu}                                  &8.48 &5.54 &10.25 &~  &1.492  \\
  &PBE-TS$^*$          &8.48 &5.53 &10.26 &~  &1.491   \\
 \hline
 \hline
$^*$ This work.
\end{tabular}
\end{table}

We also studied a set of other polymers (including PGA, PVC, POM, PPO, PPS). 
The predicted crystal structures, lattice parameters and densities for all polymers considered are shown in Table \ref{results}.
It can be clearly seen that PBE-TS significantly improves the agreement with available experimental data, compared with PBE without any vdW correction. 
The PBE-TS results obtained in our search, are slightly different from our previous study \cite{CSLiu}.
indicating a flat energy landscape of polymers. 
The deviations are mainly from the lattice vectors in non-fibre axis which are sensitive to the description of vdW dispersions.
In general, the comparison from Table \ref{results} proves that 1) TS-VDW correction allows one to reproduce the experiment lattice parameters very well, thus is sufficient to describe inter chain interactions; 2) our EA approach is very efficient for predicting polymeric structures and their crystal packing.
Moreover, such searches from small molecular building blocks can yield a comprehensive picture of the energy landscape (as shown in the examples of PA). 
Not only to identify the ground state configuration, while the low energy metastable phases can also be observed in this type of search \cite{meta}.
With the encouragement, we proceed to the cases of more complex polymers.

\begin{figure*}
\epsfig{file=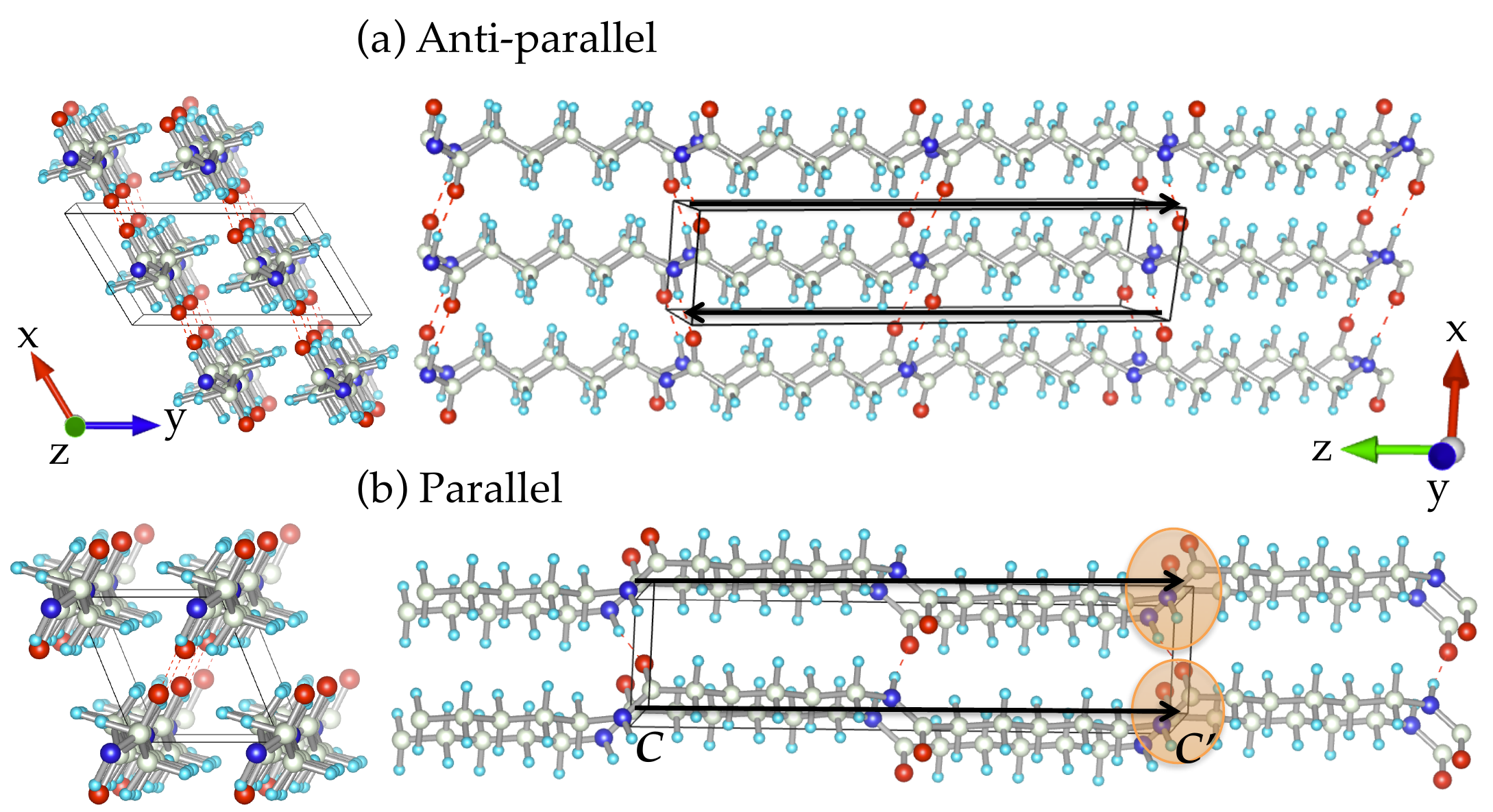, width=0.7\textwidth}
\caption{\label{Nylon} Crystal structure of Nylon 6: (a) $\gamma$ phase; (b) another low-energy configuration. Note that the two structures differ in the direction of two adjacent H-bonded sheets: anti-parallel in (a), and parallel in (b). Structure initialization in the context of line chain mode is also represented. $C$ and $C'$ are the geometric centers of monomers. The monomers are assembled in such a way that the $C$-$C'$ connections are parallel (or anti-parallel) to the $c$-axis of the cell. The unsaturated connecting groups (CO- and -NH) are marked. The degrees of freedom for each monomer include the position of its geometric center ($C$) and its rotation along $c$-axis in $ab$-plane. Grey, cyan, red and blue spheres represent the carbon, hydrogen, oxygen and nitrogen atoms, respectively.}
\end{figure*}

\begin{figure*}
\epsfig{file=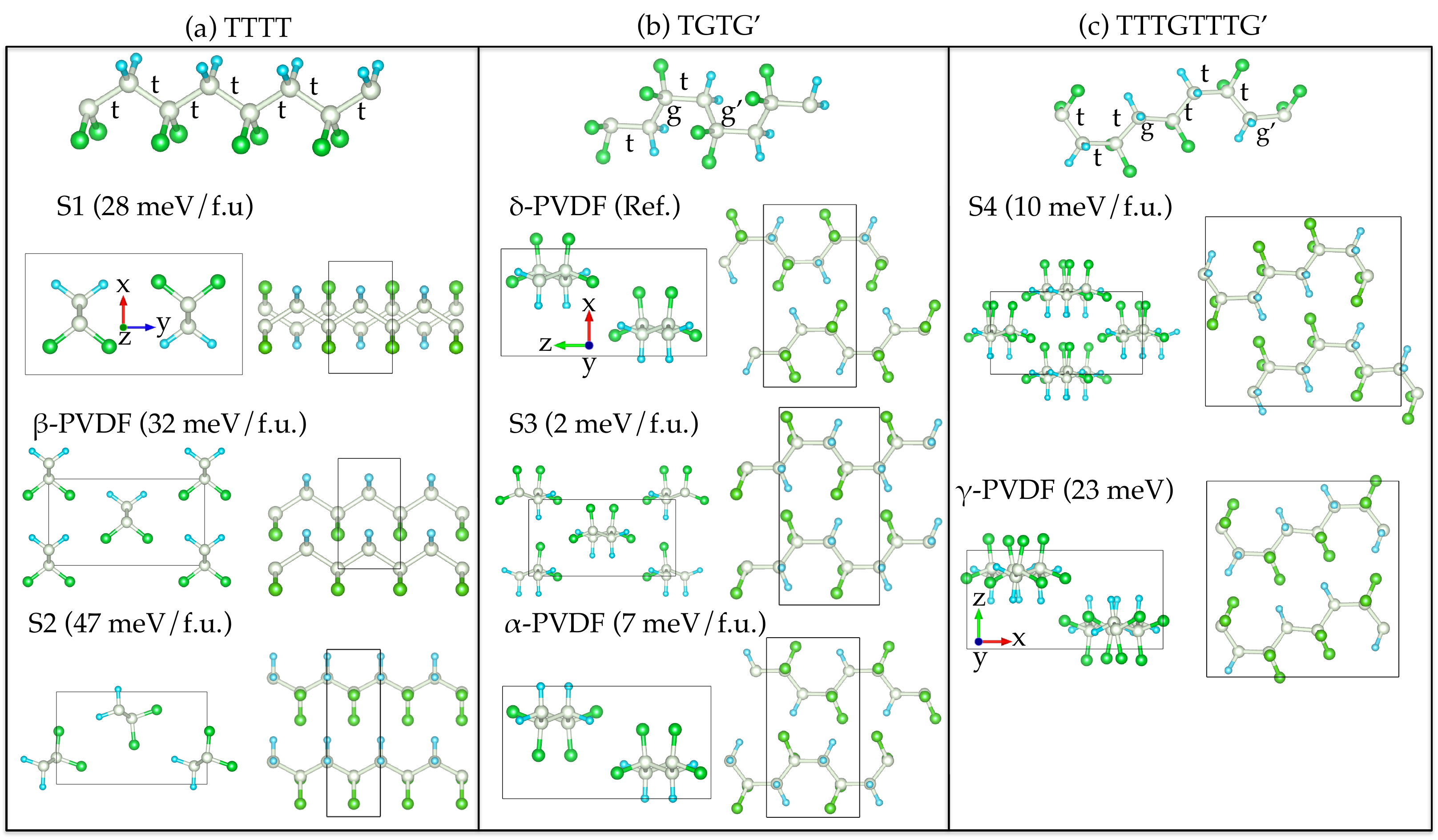, width=0.8\textwidth}
\caption{Top and side views of low-energy crystal structures of PVDF found in USPEX searches starting from (a) TTTT chain; (b) TGTG' chain; (c)TTTGTTTG' chain.  The energetics relative to the ground state ($\delta$-PVDF) are also shown. Grey, cyan and green spheres represent the carbon, hydrogen and  fluorine atoms, respectively.}
\label{PVDF}
\end{figure*}

\section{Prediction of the crystal structures of complex linear polymers}
So far, we have demonstrated a general framework to predict the crystal structures of polymers, from only the building blocks.
The prediction still needs to sample a large configurational space, including the connectivity sequence, the conformational diversity, and the manner in which the chains pack together.
This method, although very powerful in predicting polymers made of simple monomers in all trans conformation,
is very likely to face hurdles for complex polymers systems.
%One particularly important special case - that of a linear polymer - allows for further simplification, which will allow much more complex systems to be explored. 
Considering that most of the existing crystalline polymers are composed of neatly packed straight chains,
let us focus on linear polymers.  %\cite{comment2}. 
%(there might exist some crystalline branched or cross-linked polymers, but they are not the scope of this paper).
The conformation of the chain, as the primary interest in polymer chemistry, has been extensively studied.
Therefore, we can simplify the searching problem by starting from the conformation of an individual chain, and predicting the optimal pack of such chains.
%Hence there are in essential two variables to describe its crystal structure, the conformation of the chains, and their lateral packing. 

Provided that the chain conformation is known, the factors of defining their packing in the crystal are, 
1) the relative positions of the chains in the crystal; 
2) the degree of freedom of rotation associated with the lateral groups: 
3) the orientation of the chains. 
In this case, another assumption can be made:
all the infinite chains in a crystalline structure must be parallel or antiparallel to each other.
Therefore, we propose a new structure prediction scheme of {\it Linear chain mode},
where we assume the polymeric chain runs parallel to the crystallographic $c$-axis.
%For each monomer, there should be two unsaturated atoms to connect its adjacent monomer.
Here, we assemble the monomers by ensuring the neighboring contacts of these bridging atoms are close to the real situation (in terms of bond length and bond angle).
Mathematically, the propagation orientation can be determined by the vector between the 
geometric centers of two connected monomers, $C-C'$, as shown in Fig \ref{Nylon}, where structure initialization of nylon 6 in the context of linear chain mode is demonstrated.
Thus we can reorient the linear chain in the (001) or (00$\overline{1}$) direction.
To predict the crystal structures, initially we create a 2D primitive cell in the $a-b$ plane for the geometric centers, according to the randomly assigned plane group symmetry.
Then the 3D lattice is built, and the reoriented monomers around the geometric center was constructed.
During the course of new structure generation the chain orientations are randomly assigned, allowing the freedom of parallel and anti-parallel packings.
To enrich the structural diversity, a certain degree of variation from the rotation and translation of polymers along $c$-axis is permitted.
Accordingly, the rotational axis is fixed to the $c$ direction when rotational mutations are operated.

By imposing the above constraints, the linear chain mode significantly speeds up the searching process. 
Here we illustrate its power by the prediction of all poly(vinylidene fluoride) (PVDF) polymorphs, and two other well known complex polymers, $\gamma$ nylon 6 and cellulose-I$_\beta$.
\begin{table}
\caption{The predicted lattice parameters and density for all polymorphs of PVDF. For comparison available experimental and computations results at the same level of theory are also listed.}
\label{PVDF-table}
\begin{tabular}{clccccc}
\hline
\hline
Polymer & Method &\emph{a}({\AA}) &\emph{b}({\AA}) &\emph{c}({\AA}) &$\beta$($^\circ$) &Density \\
 &~&~&~&~&~& (g/cm$^3$)   \\
\hline
\multirow{3}{*}{\begin{tabular}{c}$\alpha$  \\ $Z$ = 4\\(\emph{P}2$_1$/\emph{c})  \end{tabular}}
 &Expt.\cite{Miller-1960}   &9.64  &4.96 &4.62 &90.0  &1.92     \\
 &PBE$^*$                   &9.83  &5.07 &4.68 &90.0  &1.82 \\
  &PBE-TS$^*$               &9.32  &4.83 &4.67 &90.0  &2.02   \\
 \hline
 \multirow{3}{*}{\begin{tabular}{c}$\beta$  \\$Z$ = 2\\ (\emph{Cm2m})  \end{tabular}}
 &Expt.\cite{Miller-1960}   &8.58 &4.91 &2.56 &~  &1.97   \\
  &PBE$^*$                  &8.75 &4.91 &2.58 &~  &1.91  \\
  &PBE-TS$^*$               &8.67 &4.81 &2.59 &~  &2.12   \\
\hline
  \multirow{3}{*}{\begin{tabular}{c}$\gamma$ \\ $Z$ = 8\\ (\emph{Pca}2$_1$)  \end{tabular}}
 &Expt.\cite{Lovinger40}    &4.97 &9.18 &9.66 &~  &1.93  \\
 &PBE$^*$                   &5.01 &9.32 &9.81 &~  &1.85 \\
  &PBE-TS$^*$               &4.81 &9.29 &9.55 &~  &1.99  \\
 \hline
 \multirow{3}{*}{\begin{tabular}{c}$\delta$ \\ $Z$ = 4\\ (\emph{Pna}2$_1$)  \end{tabular}}
 &Expt.\cite{Li-2013}       &4.96 &9.64 &4.62 &~  &1.93    \\
 &PBE$^*$                   &5.04 &10.01 &4.68 &~ &1.80  \\
  &PBE-TS$^*$               &4.83 &9.21 &4.63 &~  &2.04    \\
\hline
 \hline
$^*$ This work.
\end{tabular}
\end{table}

{\bf PVDF}:
is composed of polar [-CH$_2$-CF$_2$-]$_n$ repeating units.
The molecular chain can be assembled in different conformations, depending on the trans (T) or gauche (G) linkages.
The variations of chain conformation and arrangements of dipole moments lead to polymorphism.
So far, four different known phases have been well characterized experimentally; $\alpha$ (TGTG'), $\beta$ (TTTT), $\gamma$ (TTTGTTTG') and $\delta$ (TGTG') \cite{Ramer-1,Ramer-2}.
The {\it general mode}, efficiently to predict the $\beta$ phase with all-T chains, but fails to obtain other known polymorphs with complex chain conformations.
Here, we perform several searches using the new scheme of {\it linear chain mode}, by starting from chains in different conformations, namely TT, TGTG' and TTTGTTTG'.
For the TT chain, we found the $\beta$ phase, which is well known for its piezoelectric properties and is a prototype family of piezoelectric materials.
Interestingly, another two low-energy phases (S1 and S2 as shown in Fig. \ref{PVDF}) are also observed, which differ from the $\beta$ phase in the orientation of the dipole moments.
In S1 the orientations of the dipoles moments are antiparallel, while a non-collinear orientation of neighboring chain dipoles is observed in S2.
The monoclinic $\alpha$-PVDF and orthorhombic $\delta$-PVDF are successfully identified as well in the search starting from TGTG' chain.
In $\alpha$-PVDF, the dipole moments are antiparallel and mutually cancelled, while all dipole moments are oriented in the same direction in $\delta$-PVDF.
Therefore, $\alpha$ is a non-polar phase, while $\delta$ is polar.
In both phases, the adjacent chains are antiparallel.
Interestingly, we also noticed an energetically competitive configuration in TGTG' conformation with two adjacent chains being parallel (denoted as S3 in Fig. \ref{PVDF}).
Starting from the TTTGTTTG' chain, we observed that the most stable configuration (S4) consists of two parallel adjacent chains with dipoles are arranged in the same direction.
The experimentally known $\gamma$ phase, are also identified as a metastable phase in this search.
Among all of the discovered structures, $\delta$-PVDF has the lowest energy. 
However, the energy differences are notably small, which is in consistent with the previous computational studies. \cite{Goddard-PRB-2004}
The predicted lattice parameters for all known PVDF polymorphs are summarized in Table \ref{results}.
Clearly, our predictions based on PBE-TS functional are in satisfactory agreement with experiments.

{\bf $\gamma$ nylon 6}:
Two crystalline forms of nylon 6 have been experimentally well characterized, namely $\alpha$ and $\gamma$. 
The $\alpha$ phase is composed of fully extended chains, possessing 8 repeating units of [-(CH$_2$)$_5$-CO-NH-] per unit cell, while the $\gamma$ phase has simpler packing ($Z$=4). 
The crystal is composed of the pleated sheets of the parallel chains joined by hydrogen bond (the length is 1.83 \AA ~for NH--O). 
In $\gamma$ nylon 6, the chain directions are opposite in alternating sheets. 

We performed a search with $Z$ = 4.
Indeed, we found the most stable configuration is in monoclinic symmetry (space group $P$2$_1$/$a$ without considering H atoms \cite{comment1}), $a$ = 4.77 \AA, $b$ = 8.35 \AA, $c$ = 16.88 \AA (fiber axis), $\gamma$ = 121.2$^\circ$, 
in good agreement with the experimental results \cite{Nylon-Expt} except that there is a considerable deviation in cell vector $b$ (the direction between the alternating sheets with purely vdW bonding) as shown in Fig \ref{Nylon}. 
Structural topology does indeed correspond to the experimental $\gamma$ phase.
Compared with the previous study derived from classical force field \cite{Li}, the agreement has been significantly improved.
Interestingly, we also observed another extremely low-energy configuration in which the corresponding chains in adjacent sheets are parallel, with 3 meV/f.u. higher than $\gamma$ phase in energy.

\begin{figure*}
\epsfig{file=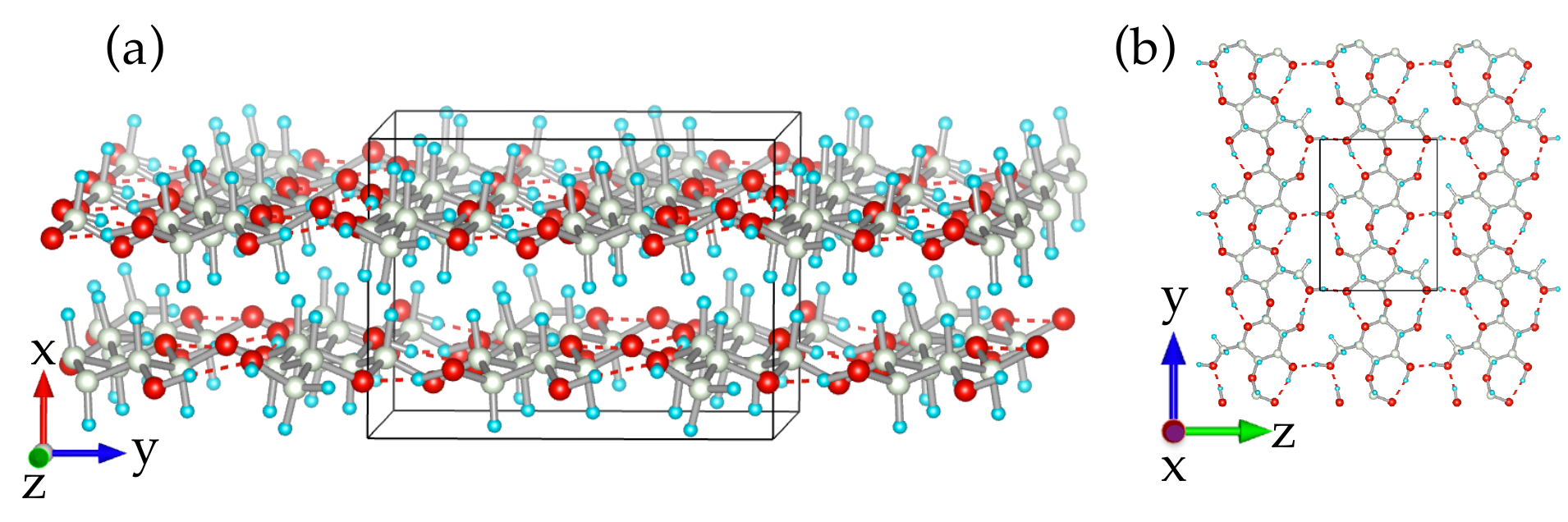, width=0.7\textwidth}
\caption{\label{cellulose} Crystal structure of Cellulose-I$_\beta$ (a) side view (b) top view. Grey, cyan and red spheres represent the carbon, hydrogen and oxygen atoms, respectively.}
\end{figure*}

{\bf Cellulose-I$_\beta$}:
Cellulose is a polymer with repeating D-glucose units [-C$_6$H$_10$O$_5$-]$_n$.
Microfibrils of naturally occurring cellulose have two crystal forms, I$_\alpha$ and I$_\beta$. 
It was found that I$_\beta$ is the thermodynamically more stable.
Although its crystal structure has been intensively studied \cite{Cellulose-Expt, Cellulose-CSP}, the crystallographic coordinates were only recently reliably determined by low temperature neutron crystallographic techniques \cite{Nishiyama-Bio-2008}. 
A very recent computational study \cite{TS-VASP} showed that TS-vdW method yields a remarkable agreement with the experimental reports, 
and we use the same choice of PBE-TS functional in our search.
Starting from the D-glucopyranosyl chains ($Z$=4), we indeed identified I$_\beta$ as the ground state configuration, 
and the calculated unit cell parameters (a=7.36 \AA, b=8.16 \AA, c=10.44 \AA, $\gamma$=96.4$^\circ$) agree well with the previous reports \cite{TS-VASP, Nishiyama-Bio-2008}(see Table \ref{table2}).
As shown in Fig. \ref{cellulose}a, the cellulose chains are arranged parallel-up and edge to edge, making flat sheets that are held together by H bonds. 
Sheets formed by H-bonded D-glucopyranosyl chains are in bc plane, while there are no strong H bonds perpendicular to the sheets. 
Most importantly, the complex hydrogen bond network in the flat sheets are also correctly recognized (Fig. \ref{cellulose}b). 

\begin{table}
\caption{\label{table2} The comparison of cell parameters for $\gamma$ nylon 6 and cellulose-I$_\beta$.}
\begin{tabular}{clccccc}
\hline
\hline
                             & Method                            & $a$ (\AA) & $b$ (\AA) & $c$ (\AA)  & $\gamma$($^\circ$)  & Density \\
                              &~&~&~&~&~& (g/cm$^3$)   \\
\hline
$\gamma$ nylon 6             & Expt.{\tiny 295 K}\cite{Nylon-Expt} &  4.78   & 9.33 &  16.88   & 121.0    &   1.17      \\
{[-(CH$_2$)$_5$-CO-NH-]$_n$} & Force field  \cite{Li}            &  4.97     & 8.68 &  17.33   & 127.5    &   1.26      \\
 $Z$ = 4                     & PBE*                              &  4.90     & 9.98 &  16.81   & 120.0    &   1.06      \\
($P$2$_1$/$a$)               & PBE-TS*                           &  4.77     & 8.35 &  16.98   & 121.2    &   1.29      \\
\hline
& & & & & &\\
cellulose-I$_\beta$          & Expt.{\tiny 15 K}\cite{Nishiyama-Bio-2008}    & 7.64   & 8.18  & 10.37    & 96.5    &   2.12        \\
{[-C$_6$H$_{10}$O$_5$-]$_n$} & PBE-TS\cite{TS-VASP}                          & 7.63   & 8.14  & 10.41    & 96.4    &   2.17        \\
$Z$ = 4                      & PBE \cite{VASP-VDW}                           & 8.70   & 8.23  & 10.46    & 95.5    &   1.87        \\
($P$2$_1$)                   & PBE-TS*                                       & 7.49   & 8.13  & 10.42    & 96.4    &   2.19        \\
\hline
\hline
$^*$ This work.
\end{tabular}
\end{table}

\section{Conclusions}
To predict crystal structures of polymers, we have developed a constrained EA operating with well-defined molecular units or blocks.   
The key feature of this approach is that each block is treated as a building block. 
At the same time, this strategy makes the problem well-defined, significantly reduces the search space and improves the efficiency of search. 
The diversity of the population of structures is enhanced by using space group symmetry combined with random cell parameters, and random positions and orientations of molecular units. 
The new constrained EA is successfully tested and validated on a diverse range of experimentally known polymers. 
By fixing the orientation of polymeric chains in the search, some complex linear polymers can be also predicted.
The excellent agreement between the predicted crystal structures and available experimental results not only elucidate the reliability of the method in the accurate prediction of the crystal structures of the polymers considered, 
but also suggests that the new method is a viable tool for the design of the future polymer materials.
For example, using this approach one can optimize not only stability, but also various other physical properties (e.g. density\cite{Zhu-PRB-2011}, hardness\cite{Lyakhov-PRB-2011}, dielectric constants\cite{Zeng-Acta-2014} etc). 

\section{acknowledgements}
We thank the National Science Foundation (EAR-1114313, DMR-1231586), DARPA (Grants No. W31P4Q1210008 and No. W31P4Q1310005), the Government (No. 14.A12.31.0003) and the Ministry of Education and Science of Russian Federation (Project No. 8512) for financial support, and Foreign Talents Introduction and Academic Exchange Program (No. B08040). Calculations were performed on XSEDE facilities and on the cluster of the Center for Functional Nanomaterials, Brookhaven National Laboratory, which is supported by the DOE-BES under contract no. DE-AC02-98CH10086. V.S. and R.R. acknowledge the Office of Naval Research for Multidisciplinary University Research Initiative (MURI) research grant.
Dr. Bucko is also acknowledged for providing the structure model of cellulose.
%\end{acknowledgement}

\bibliography{reference}

\begin{thebibliography}{49}
\bibitem [{met()}]{meta}%
  \BibitemOpen
  \href@noop {} {}\bibinfo {note} {Due to the property of EA to zoom in on the
  most promising regions of the search space - though there is no guarantee
  that all low-energy metastable states will be found by the EA.}\BibitemShut
  {Stop}%
\bibitem [{com()}]{comment1}%
  \BibitemOpen
  \href@noop {} {}\bibinfo {note} {If H atoms are considered, the resulting
  symmetry is $P$2$_1$.}\BibitemShut {Stop}%
\end{thebibliography}
\end{document}